\documentclass[12pt,showpacs,nofootinbib]{revtex4-1}
\usepackage{multirow}
\usepackage{amssymb}
\usepackage{tikz}
\usepackage{amssymb}
\usepackage{amsmath,graphicx,dsfont,mathtools}
\usepackage{physics,tensor,cancel}
\usepackage{esvect}
\usepackage{makecell}
\usepackage[compat=1.1.0]{tikz-feynman}
\usepackage{array}
\usepackage{dcolumn}
\usepackage{braket}
\usepackage{bm}
\usepackage{enumerate}
\usepackage{slashed}
\usepackage{epstopdf}
\usepackage{amsmath}
\usepackage{dcolumn,comment,caption,booktabs}
\usepackage{slashed}
\usepackage{amsfonts}
\usepackage{adjustbox}
\usepackage{tcolorbox}
\usepackage{float}
\usepackage{collectbox}
\usepackage{listings}
\usepackage[unicode]{hyperref}
\usepackage[mathscr]{eucal}
\usepackage{graphics}
\usepackage{subcaption}
\newcommand{\crb}[1]{{\color{blue}#1}}

\newcommand{\be}{\begin{equation}}
	\newcommand{\ee}{\end{equation}}
\newcommand{\bea}{\begin{eqnarray}}
	\newcommand{\eea}{\end{eqnarray}}
\newcommand{\ben}{\begin{enumerate}}
	\newcommand{\een}{\end{enumerate}}
\newcommand{\bde}{\begin{widetext}}
	\newcommand{\ede}{\end{widetext}}

\newcommand{\bc}{\begin{center}}
	\newcommand{\ec}{\end{center}}

\newcommand{\hc}{\text{h.c.}}
\newcommand{\SO}[1]{\text{SO}(#1)}
\newcommand{\SU}[1]{\text{SU}(#1)}

\begin{document}
	\newcommand{\AdrHEPC}{$^a$Department of Theoretical Physics, University of Science, Ho Chi Minh City 70000, Vietnam\\ $^b$Vietnam National University, Ho Chi Minh City 70000, Vietnam}
	
	\title{Electroweak phase transition with the confinement scale of the strong sector or dilaton in the minimal composite Higgs model}
	\author{Vo Quoc Phong$^{a,b}$}
	\email{vqphong@hcmus.edu.vn}
	\affiliation{\AdrHEPC}
	\author{Truong Van Tien$^{a,b}$}
	\email{tvtien832@gmail.com}
	\affiliation{\AdrHEPC}
	\author{Phan Hong Khiem$^{c,d}$}
	\email{phanhongkhiem@duytan.edu.vn}
	\affiliation{$^c$Institute of Fundamental and Applied Sciences, Duy Tan University, Ho Chi Minh City 70000, Vietnam\\
		$^d$ Faculty of Natural Sciences, Duy Tan University, Da Nang City 50000, Vietnam}	
	\begin{abstract}
The minimal Composite Higgs model (MCHM) provides an effective trigger for the Baryogenesis scenario through the confinement scale of the strong sector ($f$) or dilaton ($\chi$). $f$ is a parameter with mass dimension, which stores the resonances of particles at high energies and has a suitable value of about $800$ GeV. But when $300$ GeV $\le f \le 400$ GeV, the effective Higgs potential has a first-order electroweak phase transition. Therefore, although $f$ cannot be a perfect trigger, it does suggest an effective approach that accommodates the resonances of particles. Thus the investigation of the electroweak phase transition according to $f$ has confirmed that the inclusion of the dilaton in the effective potential is reasonable. Accordingly, we derive a dilaton potential with appropriate parameter domains and $f=800$ GeV; the mass of the dilaton ranges from $300$ GeV to $700$ GeV, which will give an electroweak phase transition strength greater than $1$ and less than $3$, enough for a first-order phase transition. This is a direct and clear evidence of the triggers for the first-order EWPT in the MCHM.
\end{abstract}
	\maketitle
\tableofcontents
\section{INTRODUCTION}\label{secInt}
	
The baryogenesis scenario associated with Sakharov's conditions \cite{sakharov}, is analyzed in a wide variety of models with different triggers \cite{plv,2b,2c,BSM,BSMb,majorana,majoranab,thdmb,ESMCO,elptdm,elptdma,elptdmb,elptdmc,elptdmd,phonglongvan,phonglongvanb,phonglongvan2,SMS,munusm,lr,singlet,singletb,singletc,singletd,mssm1,mssm1b,mssm1c,twostep,twostepb,twostepc,1101.4665,1101.4665b,1101.4665c,jjgb,jjgc,jjgd,Ahriche1,Ahriche2,Ahriche2b,Ahriche3,span,chr,cde,kusenko}. Most of them are associated with a new physical prediction such as an energy scale, a dark matter candidate \cite{plv,phonglongvan2,ptl} or a change in some coupling in the SM \cite{yukawa1,yukawa2}.

The Composite Higgs model \cite{dugan} is also a possible solution to this problem. The model argues that the SM is in fact an effective theory and the influence of very high energy physics on the current effective theory at low energies through the condensation of strongly bound states before the electroweak symmetry breaking occurs and can spontaneously trigger the EWPT. In addition to the EWPT, the model helps explain the origin of the Higgs particle, which is actually formed from the symmetry breaking of a larger group at high energies.

However, there is no basis for determining the high-energy symmetry group, which is the core of the Composite Higgs Model. The Composite Higgs Model (CHM) borrows ideas from the techinicolor model to explain the origin of the Higgs particle as well as the complex doublet $\phi$ that appears in the Standard Model Lagrangian. The CHM suggests that the complex doublet is formed by Nambu-Goldstone bosons (NGBs) arising from the breaking of the symmetry group $\mathcal G$ to the symmetry group $\mathcal H$ such that the electroweak group $\mathcal G_{EW}\subset\mathcal H$, and the Higgs particle is a pseudo-NGB (pNGB) among these NGBs. The non-SM fields that appear at high energies will become effective interactions between elementary particles at low energies, and the SM is now an effective theory. Therefore, the Lagrangian of the CHM now needs to be modified due to the influence of non-fundamental fields, also known as CFT fields. The part of the Lagrangian related to these fields is called the composite part. In the scope of this paper, we choose the minimal Composite Higgs model (MCHM) \cite{agashe_minimal_2005} for our investigation, in which the Higgs doublet is parameterized in the \SO{5}/\SO{4} coset.

Therefore, many Composite Higgs models have been studied, many of which are still insufficient to trigger electroweak symmetry breaking \cite{fujikura}. Besides the efforts to find a suitable symmetry group, it has been theorized that it is possible to modify the Composite Higgs model by "adding" the dilaton field \cite{bruggissera, bruggisserb}. 

First of all, we can make a general observation that in the MCHM, the meson resonance components (stored in the confinement scale of the strong sector ($f$) or the parameters of effective potential) and dilaton will push the effective potential differently from the SM and make the phase transition more violent than the SM. This has been pointed out in Refs. \cite{bruggissera,bruggisserb}. But with a large number of parameters, specifying a first-order EWPT is quite difficult. Therefore, we will conduct the survey of the EWPT process using a method tentatively called like-SM in Sec.\ref{iii}.
	
The paper has the following structure. Except for the Introduction (Sec.~\ref{secInt}) and the Conclusion and Outlooks (Sec.~\ref{vi}), the paper quantitatively demonstrates the EWPT process with a one-loop effective potential. Sec.~\ref{iii} gives a quick review of the CHM, as well as comments and remarks on the masses of particles in the CHM. In Sec.~\ref{iv}, the effective potential will be studied; then, the EWPT strength will be calculated for different values of $f$. In Sec.~\ref{v}, the effective Higgs-dilaton potential is used to recalculate the electroweak phase transition strength $(S)$ for a mass range of the dilaton. $S$ is the phase transition strength, which is the ratio between the minimum of the effective potential and the critical temperature.
	
\section{Review on the minimal composite Higgs model}\label{iii}

Unlike the SM, where the Higgs doublet is initially unknown, in the CHM it is created by another symmetry breaking. Therefore, when studying the CHM, we need to specifically determine the vacuum state to analyze the breaking of $\mathcal G\to\mathcal H$ before the electroweak symmetry breaking occurs.

In general, the group $\mathcal G$ has a Lie group algebra generated by generators $T^A$. $T^A$ is divided into two parts: the unbroken generators $T^a$ ($a=1...\dim[\mathcal H]$) are the Lie algebraic basis of the group $\mathcal H$, and the broken generators $\widehat T^{\hat a}$ ($\hat a=1...\dim[\mathcal G/\mathcal H]$). From this, a vacuum configuration of the composite part $\vv F$ can be chosen such that
\begin{equation}
	T^a\vv F=\vv0,\quad\widehat T^{\hat a}\vv F\neq \vv0.	
\end{equation}
The NGBs $\theta^{\hat a}$ are parameterized as a local transformation in the direction of $\widehat T^{\hat a}$ generators:
\begin{equation}
	\vv\Phi(x)=\exp[i\theta^{\hat a}(x)\widehat T^{\hat a}]\vv F.
\end{equation}
Then, if $\theta$ reaches VEV $\ev{\theta}\neq0$, it will cause the breaking of $\mathcal G_{EW}$ symmetry in $\mathcal H$. Geometrically, $\ev{\theta}$ is the angle between $\vv F$ and $\vv\Phi$. The breaking of $\mathcal G_{EW}$ can be represented by the "projection" of $\vv\Phi$ onto $\mathcal H$ being nonzero after applying a rotation onto $\vv F$ (which was initially "perpendicular" to $\mathcal H$ due to $T^a\vv F=0$), i.e.,
\begin{equation}
	v=f\sin\ev\theta\neq 0,
\end{equation}
with $f=|\vv F|$. $v$ is the VEV that causes the electroweak symmetry breaking. This is called the vacuum difference, i.e., the difference between the VEV of the  $\mathcal G\to\mathcal H$ $f$ breaking and the VEV of the electroweak symmetry breaking $(v)$. And unlike the technicolor model, the electroweak symmetry breaking in the CHM occurs if and only if the angle is very small,
\begin{equation}
	\epsilon\equiv\frac{v^2}{f^2}=\sin^2\ev\theta\ll1.\label{4}
\end{equation}
Depending on how the model is constructed, the value of $\epsilon$ can be obtained naturally, or it can be calibrated to a greater or lesser extent.

To construct the CHM $\mathcal G\to\mathcal H$, where the electroweak group $G_{EW}=SU(2)\times U(1)\subseteq\mathcal H$, we need a complex Higgs doublet for it, i.e., at least 4 Goldstone bosons. To find the $\mathcal G$ and $\mathcal H$ groups, one way is to consider the breaking of the SO$(N)$ group. By the Goldstone's theorem, the spontaneous symmetry breaking \mbox{SO$(N)\to$ SO$(N-1)$} will generate $N-1$ Goldstone bosons. Thus, the smallest possible case of $N$ is \mbox{SO(5) $\to$ SO(4)}. The 10 generators of the group $SO(5)$ in the basic representation, $T^A=\{T^a,\widehat T^{\widehat a}\}$, are divided into two parts as follows:
\begin{gather}
	T^a=\qty{T^\alpha_L=\mqty[t^\alpha_L&\vv0\\[0.5em]\vv0^T&0],\ T^\alpha_R=\mqty[t^\alpha_R&\vv0\\[0.5em]\vv0^T&0]}\quad({\alpha=1,2,3}),\\
	(\widehat{T}^{\hat a})_{IJ}=-\frac{i}{\sqrt{2}}(\delta^i_I\delta^5_J-\delta^i_J
	\delta^5_I).
\end{gather}
The 6 generators $T^a$ generate the group algebra SO(4) according to the generators of SU$(2)_L\times$ SU$(2)_R$ because SO(4) is isomorphic to the chiral group SU$(2)_L\times$ SU$(2)_R$
\begin{equation}
	\begin{gathered}
		(t^\alpha_L)_{ij}=\frac{1}{4}\Tr[\overline\sigma_i^\dagger\sigma^\alpha\overline\sigma_j]=-\frac{i}{2}\qty[\varepsilon_{\alpha\beta\gamma}\delta^\beta_i\delta^\gamma_j+(\delta^\alpha_i\delta^4_j-\delta^\alpha_j\delta^4_i)],\\
		(t^\alpha_R)_{ij}=\frac{1}{4}\Tr[\overline\sigma_i\sigma^\alpha\overline\sigma_j^\dagger]=-\frac{i}{2}\qty[\varepsilon_{\alpha\beta\gamma}\delta^\beta_i\delta^\gamma_j-(\delta^\alpha_i\delta^4_j-\delta^\alpha_j\delta^4_i)].
	\end{gathered}
\end{equation}
We parameterize the 4 NGBs $\Pi^{\hat a}$ corresponding to the 4 generators $\widehat T^{\widehat a}$ broken into
\begin{equation}
	\vv\Phi=\exp[i\frac{\sqrt2}{f}\Pi^{\hat a}\widehat T^{\hat a}]\vv F\equiv U[\vv\Pi]\vv F.
\end{equation} 
In which, $U[\vv\Pi]$ is called the Goldstone matrix, which can be explicitly calculated as:
\begin{equation}
	U[\vv\Pi]=\mqty[I_4-\qty(1-\cos\dfrac{\abs{\Pi}}{f})\dfrac{\vv\Pi\vv\Pi^T}{\abs{\Pi}^2}&\sin\dfrac{\abs{\Pi}}{f}\dfrac{\vv\Pi}{\abs{\Pi}}\\[1em]-\sin\dfrac{\abs{\Pi}}{f}\dfrac{\vv\Pi^T}{\abs{\Pi}}&\cos\dfrac{\abs{\Pi}}{f}],
\end{equation}
where $\abs{\Pi}=\sqrt{\vv\Pi^T\vv\Pi}$. 

\begin{equation}
	\vv F=\mqty[\vv 0\\f]\Rightarrow\vv\Phi=f\mqty[\sin\dfrac{\abs{\Pi}}{f}\dfrac{\vv\Pi}{\abs{\Pi}}\\\cos\dfrac{\abs{\Pi}}{f}],
\end{equation}
with $f=\abs{\vv F}$. As said, 4 $\Pi^{\hat a}$ can form a 1/2 hypercharged Higgs doublet:
\begin{equation}
	\phi=\mqty[\phi^+\\\phi^0]=\frac{1}{\sqrt2}\mqty[\Pi^2+i\Pi^1\\\Pi^4-i\Pi^3]\Leftrightarrow\vv\Pi=\mqty[\Pi^1\\\Pi^2\\\Pi^3\\\Pi^4]=\frac{1}{\sqrt2}\mqty[-i\qty(\phi^+-{\phi^+}^\dagger)\\\phi^++{\phi^+}^\dagger\\i\qty(\phi^0-{\phi^0}^\dagger)\\\phi^0+{\phi^0}^\dagger].
\end{equation}
This doublet $\phi$ is the same one mentioned in the SM. And similar to the SM, we can choose the unitary gauge for $\vv\Pi$
\begin{equation}
	\phi=\mqty[0\\H],\vv\Pi=\mqty[0\\0\\H\\0]\Rightarrow U[H]=\mqty[I_3&\vv0&\vv0\\\vv0^T&c_H&s_H\\\vv0^T&-s_H&c_H]
	\Rightarrow\vv\Phi=\mqty[\vv0\\fs_H\\fc_H],\label{eq:unitaPhi}
\end{equation}
with the field $H$ considered in the perturbation of VEV, $\ev{H}=h$, i.e. $H(x)=h+\delta h(x)$ as in the SM. The trigonometric functions have been re-noted for brevity: $s_H\equiv\sin(H/f),\ c_H\equiv\cos(H/f)$.

However, to be able to construct fermion multiplets in the MCHM, an additional group $U(1)_X$ is needed, i.e.,
\begin{equation}
	\text{SO(5)}\otimes\text{U(1)}_X\to\text{SO(4)}\otimes\text{U(1)}_X.
\end{equation}
We will embed the gauge field $X_\mu$ of U$(1)_X$ into the field $B_\mu$ by setting $X_\mu=B_\mu$ with the coupling constant also being $g'$. Then the hypercharge is redefined as $Y=T^3_R+X$. Each quark will have its own $X$ to give the corresponding hypercharge.

To study the electroweak breaking in the MCHM, we need to write the effective Lagrangian before the electroweak symmetry  breaking. The Lagrangian includes the composite part in the near-vacuum state and the integration of resonance fields; these fields are not in the SM and are therefore outside the Lagrangian. Specifically, we divide the Lagrangian into the following parts:
\begin{equation}
	\mathcal L=\mathcal L_\Phi+\mathcal L_g+\mathcal L_f-V_\text{eff},
\end{equation}
with the terms being the NGB part, the gauge boson part, the fermion part (top quark) and finally the effective potential energy. These components will be presented in the next section. We do not present the $\mathcal L_g$ component and can refer to Ref.\cite{agashe_minimal_2005} because it is not related to the problem under consideration.

\subsection{Spectra of particles}

The NGB part includes the kinetic energy of the NGB parameterized by the scalar field $\vv\Phi$ mentioned earlier, using matrix form,
\begin{equation}
	\mathcal L_\Phi=\frac{1}{2}(D_\mu\vv\Phi^T)(D^\mu\vv\Phi),
\end{equation}
with
\begin{equation}
	D_\mu=\partial_\mu-igW_\mu^\alpha T^\alpha_L-ig'B_\mu T^3_R.
\end{equation}
Here, if we consider $\vv\Pi$ in the perturbation of VEV, the boson mass term is also derived similarly to the SM,
\begin{align}
	\begin{aligned}[b]
		D_\mu\vv\Phi&=\partial_\mu\vv\Phi-igW^\alpha_\mu T^\alpha_L\vv\Phi-ig'B_\mu T^3_R\vv\Phi\\
		&=\partial_\mu H\mqty[\vv 0\\c_H\\-s_H]-igW^\alpha_\mu\mqty[t^\alpha_L&\vv0\\\vv0^T&0]\mqty[\vv 0\\fs_H\\fc_H]-ig'B_\mu\mqty[t^3_R&\vv0\\\vv0^T&0]\mqty[\vv 0\\fs_H\\fc_H]\\
		&=\partial_\mu H\mqty[\vv 0\\c_H\\-s_H]-igf\sin\frac{H}{f}\qty(-\frac{i}{2})\mqty[W^1_\mu\\W^2_\mu\\W^3_\mu\\0\\0]-ig'B_\mu f\sin\frac{H}{f}\qty(\frac{-i}{2})\mqty[0\\0\\-1\\0\\0].
	\end{aligned}
\end{align}
It will lead to
\begin{align}
	\begin{aligned}[b]
		\frac{1}{2}\abs{D_\mu\vv\Phi}^2&=\frac{1}{2}\abs{\partial_\mu H}^2+\frac{f^2}{8}\sin^2\frac{H}{f}\qty[g^2|W^1_\mu|^2+g^2|W^2_\mu|^2+|gW^3_\mu-g'B_\mu|^2]+\dots\\
		&\approx\frac{1}{2}\abs{\partial_\mu \delta h}^2+\frac{f^2}{8}\sin^2\frac{h}{f}\qty[g^2|W^+_\mu|^2+g^2|W^-_\mu|^2+(g^2+g'^2)|Z^0_\mu|^2]+\dots\\
	\end{aligned}
\end{align}

If $\sin[2](h/f)=v/f$, the masses of $W^\pm$ and $Z$ bosons are generated exactly as in the SM. The mass functions of bosons are different from those in the SM:
\begin{gather}
	M_W^2=\frac{g^2}{4}f^2s_h^2,\quad M_Z^2=\frac{g^2+g'^2}{4}f^2s_h^2.	
\end{gather}

Similar to the SM, we consider only the contribution of top quark to be significant. In the UV, top quark interacts with the composite fields in the form of partial composites,
\begin{equation}
	\mathcal L^\text{UV}_f=\mathcal L_f^\text{kin}+\sum_r\lambda_r\bar\psi\mathcal O_r+\hc
\end{equation}
The top and bottom quarks need to be embedded in $\bar\psi$ with a suitable representation of SO(5) so that after breaking up they give the mass and interaction terms as SM. There are many ways to choose the representation for the multiplet, and each will give different results. The simplest is to use the $\vb 5$ representation for each multiplet, namely
\begin{equation}
	Q_L=\frac{1}{\sqrt2}\mqty[-ib_L\\-b_L\\-it_L\\t_L\\0],\quad T_R=\mqty[0\\0\\0\\0\\t_R].
\end{equation}
In this case, $t_L$ is embedded in the $SO(4)$ quadruplet in the $Q_L$ multiplet as above to be exactly equivalent to the $\vb 2_{-1/2}$ doublet. Note that the zero elements in the above multiplets are actually physical values of the non-dynamical fields, and since they have no physical role, we can safely set them to zero and continue the calculation. To be able to write the Lagrangian when the composite part is near the vacuum $\vv F$, we apply a Goldstone matrix to $Q_L$ and $T_R$. We can do this because the quark multiplets have a global symmetry of $\SO{5}$ and the Goldstone matrix itself is a certain rotation of the $\SO{5}$ group,
\begin{gather}
	Q_L\to U^{-1}Q_L\Rightarrow \overline Q_L\to\overline Q_L  U=\frac{1}{\sqrt2}\bar t_L\mqty[0&0&i&c_H&s_H],\\
	T_R\to U^{-1}T_R\Rightarrow \overline T_R\to\overline T_R  U=\bar t_R\mqty[0&0&0&-s_H&c_H].
\end{gather}
After the transformation, we can write the effective Lagrangian of the fermion part in the MCHM,
\begin{equation}
	\begin{aligned}[b]
		\mathcal L_\text{mix}^\text{eff}&=i\overline q_L\gamma^\mu D_\mu q_L+i\overline t_R\gamma^\mu D_\mu t_R\\
		&\quad+\sum^{N_S}_{i=1}\overline\psi^i_1\qty(i\gamma^\mu D_\mu-m_{1i})\psi^i_1+\sum^{N_Q}_{i=1}\overline\psi^i_4\qty(i\gamma^\mu D_\mu-m_{4i})\psi^i_4\\
		&\quad+\sum^{N_S}_{i=1}\qty[y^i_{L1}f\qty(\overline Q_LU)_1\psi^i_1+y^i_{R1}f\qty(\overline T_RU)_1\psi^i_1+\text{h.c.}]\\
		&\quad+\sum^{N_Q}_{i=1}\qty[y^i_{L4}f\qty(\overline Q_LU)_4\psi^i_4+y^i_{R4}f\qty(\overline T_RU)_4\psi^i_4+\text{h.c.}].
	\end{aligned}
\end{equation}
In which, $\qty(\overline Q_LU)_{1,4}$ and $\qty(\overline T_RU)_{1,4}$ are the singlet and quadruplet in the pentaplet
\begin{equation}
	\overline Q_LU\equiv\mqty[\qty(\overline Q_LU)_{4}&\qty(\overline Q_LU)_{1}],\quad\overline T_RU\equiv\mqty[\qty(\overline T_RU)_{4}&\qty(\overline T_RU)_{1}].
\end{equation}
$\psi_{1,4}$ are also the corresponding singlets and quadruplet generated by the $\mathcal O$ operator. If we integrate the above Lagrangian out of the resonance fields, we get the effective Lagrangian of the top quark of the form
\begin{equation}\label{eq:L_eff_topquark}
	\mathcal L^\text{eff}_f=\overline t_L\cancel p\Pi_L(p^2)t_L+\overline t_R\cancel p\Pi_R(p^2)t_R-\overline t_L\Pi_{LR}(p^2)t_R-\overline t_R\Pi_{RL}(p^2)t_L.
\end{equation}
In which
\begin{align}
	&\begin{aligned}[b]
		\Pi_L(p^2)&=1-\frac{f^2}{2}s_H^2\sum^{N_S}_{i=1}\frac{\abs{y^i_{L1}}^2}{p^2-m_{1i}^2}-\frac{f^2}{2}\sum^{N_Q}_{i=1}\frac{\abs{y^i_{L4}}^2+c_H^2\abs{y^i_{L4}}^2}{p^2-m_{4i}^2}\\
		&=1-f^2\sum^{N_Q}_{i=1}\frac{\abs{y^i_{L4}}^2}{p^2-m^2_{4i}}+s^2_H\frac{f^2}{2}\qty[-\sum^{N_S}_{i=1}\frac{\abs{y^i_{L1}}^2}{p^2-m_{1i}^2}+\sum^{N_Q}_{i=1}\frac{\abs{y^i_{L4}}^2}{p^2-m_{4i}^2}]\\
		&\equiv\Pi_L^0(p^2)+s^2_h\Pi_L^1(p^2),
	\end{aligned}\label{eq:piL}\\
	&\begin{aligned}[b]
		\Pi_R(p^2)&=1-f^2c_H^2\sum^{N_S}_{i=1}\frac{\abs{y^i_{R1}}^2}{p^2-m_{1i}^2}-f^2s_H^2\sum^{N_Q}_{i=1}\frac{\abs{y^i_{R4}}^2}{p^2-m^2_{4i}}\\
		&=1-f^2\sum^{N_S}_{i=1}\frac{\abs{y^i_{R1}}^2}{p^2-m_{1i}^2}+s^2_Hf^2\qty[\sum^{N_S}_{i=1}\frac{\abs{y^i_{R1}}^2}{p^2-m_{1i}^2}-\sum^{N_Q}_{i=1}\frac{\abs{y^i_{R4}}^2}{p^2-m_{4i}^2}]\\
		&\equiv\Pi^0_R(p^2)+s^2_H\Pi^1_R(p^2),
	\end{aligned}\label{eq:piR}\\
	&\begin{aligned}[b]
		\Pi_{LR}(p^2)&=\frac{f^2}{\sqrt2}s_Hc_H\sum^{N_S}_{i=1}y^i_{L1}y^{i*}_{R1}\frac{m_{1i}}{p^2-m_{1i}^2}-\frac{f^2}{\sqrt2}s_Hc_H\sum^{N_Q}_{i=1}y^i_{L4}y^{i*}_{R4}\frac{m_{4i}}{p^2-m_{1i}^2}\\
		&=\frac{f^2}{\sqrt2}s_Hc_H\qty[\sum^{N_S}_{i=1}y^i_{L1}y^{i*}_{R1}\frac{m_{1i}}{p^2-m_{1i}^2}-\sum^{N_Q}_{i=1}y^i_{L4}y^{i*}_{R4}\frac{m_{4i}}{p^2-m_{1i}^2}]\\
		&\equiv s_Hc_H\Pi_{LR}^0(p^2),
	\end{aligned}\label{eq:piLR}\\
	&\Pi_{RL}(p^2)=\Pi^*_{LR}(p^2)=s_Hc_H\Pi^{0*}_{LR}(p^2)\equiv s_Hc_H\Pi^0_{RL}(p^2)\label{eq:piRL}.
\end{align}
The functions $\Pi^{0,1}_{L,R}$ and $\Pi^0_{LR,RL}$ need to return values that are true for the SM when considering the low-energy limit $p^2\approx 0$. Specifically,
\begin{gather}
	Z_{L,R}=\Pi^0_{L,R}(0)+\Pi^1_{L,R}(0)\epsilon,\\ m_t^2=\frac{\abs{\Pi_{LR}(0)}^2}{Z_LZ_R}=\abs{\Pi^0_{LR}(0)}^2\frac{\epsilon(1-\epsilon)}{Z_LZ_R}\approx\frac{\abs{\Pi^0_{LR}(0)}^2}{\Pi^0_L(0)\Pi^0_R(0)}\epsilon(1-\epsilon).
\end{gather}
Thus approximately, the mass of top quark has the form
\begin{equation}
	M_t^2=\frac{\abs{\Pi^0_{LR}(0)}^2}{\Pi^0_L(0)\Pi^0_R(0)}s_h^2c_h^2=\frac{m_t^2}{\epsilon(1-\epsilon)}s_h^2c_h^2=\frac{y_t^2f^2}{2(1-\epsilon)}s_h^2c_h^2.
\end{equation}
\subsection{The effective Higgs potential}

The effective Higgs potential \cite{bruggisserc}
\begin{equation}
	V_0\qty(h)=\alpha^0s_h^2+\beta^0 s_h^4; (s_h\equiv\sin(h/f),\ c_h\equiv\cos(h/f)),\label{34}
\end{equation}

\begin{equation}
	\begin{cases}
		\pdv*{V}{h}\eval_{h=\ev{h}}=0.\\
		\pdv*[2]{V}{h}\eval_{h =\ev{h}}=m_h^2.
	\end{cases}
	\Rightarrow
	\begin{cases}
		\alpha^0=-2\beta^0\epsilon,\\
		m_h^2f^2=8\beta^0\epsilon(1-\epsilon),
	\end{cases}
\end{equation}
and the minimum conditions of the potential,
\begin{equation}\label{eq:MCHM_eff}
	V_0\qty(h)=\beta^0\qty(s_h^4-2\epsilon s_h^2)=\frac{m_h^2f^2}{8\epsilon(1-\epsilon)}(s_h^4-2\epsilon s_h^2).
\end{equation}
There are two aspects of the effective potential: (I) we need to ensure $\abs{\alpha^0}\ll\abs{\beta^0}$ due to the condition $\epsilon=-\alpha^0/2\beta^0\ll1$; (II) $\beta^0>0$ for the mass term to be positive, which means $\alpha^0<0$. Depending on how the CHM is constructed, we can adjust $\alpha^0$ and $\beta^0$ so that the potential energy has the correct minimum and the Higgs mass is correct through the contributions of the additional resonances introduced. Whether the adjustment is large or not depends on the representation of fermion multiplets.

\begin{equation}
	\begin{gathered}
		M_h^2=\frac{m_h^2f^2}{8\epsilon(1-\epsilon)}\frac{\partial^2 (s_h^4-2\epsilon s_h^2) }{\partial^2 h} =\frac{m_h^2}{2\epsilon(1-\epsilon)}(3s_h^2-4s_h^4-\epsilon+2\epsilon s_h^2).
	\end{gathered}
\end{equation}

The effective Higgs potential is of the form Eq. (\ref{34}), we do not yet see the explicit contributions of other particles to this effective potential. Calculating their contributions is very complicated. The one-loop contributions of gauge bosons, top quark, and Higgs fields come from the resonance meson components. There is a less dynamical alternative: We consider the particles as in the SM, and their contributions are calculated as in the SM, but now the masses of particles are functions of $\sin(h/f)$. This is the like-SM method as introduced in  Sec.\ref{secInt}. Therefore, the effective potential at 0K and high temperatures takes the following forms,

\begin{equation}
	V_\text{eff}(h)=V_0(h)+\sum_{i=h,W,Z,t}\frac{n_i}{64\pi^2}\qty[M_i^4\qty(\log\frac{M_i^2}{m_i^2}-\frac{3}{2})+2M_i^2m_i^2].
\end{equation}

\begin{equation}\label{eq:MCHM_full_V_eff}
	\begin{aligned}[b]
		V_\text{eff}^\beta(h)&=V_\text{eff}(h)+\Delta V_1^\beta(h)\\
		&=V_\text{eff}(h)+\sum_{i=h,W,Z}\frac{n_i}{2\pi\beta^4}J_B\qty(\beta^2M_i^2)-\frac{12}{2\pi\beta^4}J_F\qty(\beta^2M_t^2),
	\end{aligned}
\end{equation}
in which
\begin{equation}
J_k(m^2\beta^2) = \int\limits_0^\infty dx\ x^2 \ln\left[1+n_k e^{-\sqrt{x^2+\beta^2m^2(\phi_c)}}\right], \quad\text{ with }n_F = 1, n_B = -1, \beta=1/T \, .
\end{equation}

The mass formulas of particles are functions that depend on $s_h$. Also note that, in the below expressions, $m_h, m_Z, m_W, m_t$ are the masses of Higgs, $W$, $Z$ bosons, and top quark at 0K, respectively,

\begin{equation}
	\begin{gathered}
		M_h^2=\frac{m_h^2f^2}{8\epsilon(1-\epsilon)}\pdv[2]{h}(s_h^4-2\epsilon s_h^2)=\frac{m_h^2}{2\epsilon(1-\epsilon)}(3s_h^2-4s_h^4-\epsilon+2\epsilon s_h^2),\\
		M^2_{W,Z}=\frac{m_{W,Z}^2}{\epsilon}s_h^2,\quad M^2_t=\frac{m_t^2}{\epsilon(1-\epsilon)}(s_h^2-s_h^4).
	\end{gathered}
\end{equation}

Therefore, the meson resonance components are stored in the parameter $\epsilon, f$ and change the mass spectral formula of the particles compared to the SM.

\begin{figure}[h]
	\centering
	\includegraphics[width=0.8\linewidth]{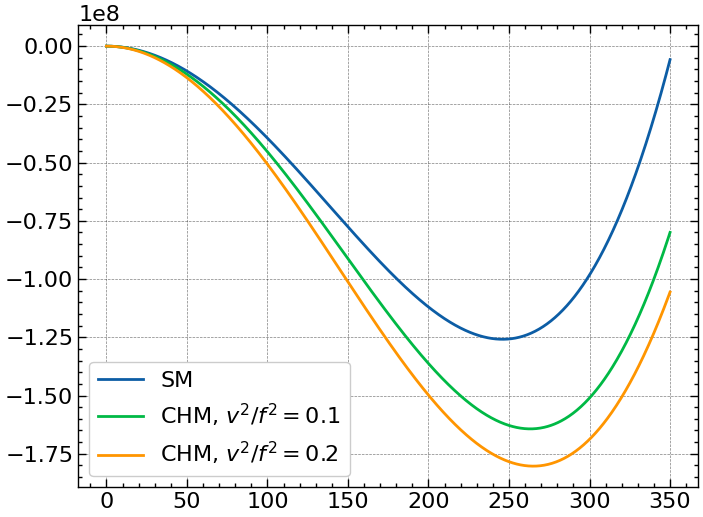}
	\caption{Plots of the effective potential at 0K in SM and MCHM for two different $\epsilon$ values.}
	\label{fig:SMCHM}
\end{figure}

Experimental results constrain that $\epsilon \leq 0.2$ \cite{17} so we choose $\epsilon=0.1$ and $\epsilon=0.2$. As shown in Fig.\ref{fig:SMCHM}, the effective potential of the MCHM has a lower minimum than that of the SM, meaning that the vacuum in the MCHM is more stable than that in SM. Due to the effect of the $\sin$ function, the VEV of Higgs is larger than 246 GeV. In addition, the larger $f$ is, the closer the effective potential of the MCHM is to that of the SM.

Another way to sum all the contributions of the SM particles is as in Ref.\cite{Ben}. Also, in Refs. \cite{bruggissera, bruggisserb, bruggisserc}, the process of calculating the effective potential is similar to ours when the high temperature components are like the last two components in Eq.(\ref{eq:MCHM_full_V_eff}). However, the authors consider the 0K components quite simply by including the contributions of the particles that are stored in the coefficients of the potential. This is also qualitatively reasonable, since the contributions of the particles at 0K do not have the trigger value for a first-order phase transition. But to calculate the phase transition strength more accurately, we have done as above way with Eq.(\ref{eq:MCHM_full_V_eff}).

\section{EWPT with the confinement scale of the strong sector}\label{iv}

With the effective potential Eq.(\ref{eq:MCHM_full_V_eff}), we proceed to find the phase transition temperatures $T_c$ and $h_m(T_c)$. $h_m(T_c)$ is the minimum of the effective Higgs potential, which represents the exact VEV causing symmetry breaking at $T_c$. Then, calculate the phase transition strength $S=\frac{h_m(T_c)}{T_c}$. The process can be solved numerically by running Eq.(\ref{eq:MCHM_full_V_eff}) at different temperatures and finding their minima.

So we have a pre-selected input parameter $f$, then run the effective potential, with temperature and output the minimum value. The temperature and the minimum value that cancel the potential, are $T_c$ and $h_m(T_c)$ respectively. Since $v=246$ GeV, instead of choosing $f$ we can choose $\epsilon$.

A strong first-order phase transition is usually understood as $S>1$. This comes from the sphaleron decoupling condition \cite{arnold,quiross,moores} and analysis of the effective potential barrier (there exists a sufficiently large maximum between two minima such that the transition from zero to nonzero minimum is violent). This condition is essentially the most important condition in the Baryogenesis scenario. The condition is not necessarily $S>1$, it can be slightly less than 1 for a first-order phase transition. However, for a strong first-order phase transition, this condition is often accepted because it involves calculating the sphaleron energy. Most models use this condition, such as the Two-Higgs-Doublet Model \cite{Baslers,ps3s}. Hence, we expect an outcome $S>1$ in this model.

As mentioned earlier, $\epsilon\lesssim 0.2$, corresponding to $f\gtrsim 550$ GeV, is of particular interest. Using the effective potential to probe the transition temperature at some values of $f$ between $300$ and $800$ GeV (corresponding to $\epsilon$ from $0.09$ to $0.24$), \crb{we obtain} the results in \autoref{tab:EWPT_CHM_500-800} and \autoref{tab:EWPT_CHM_300-500}.

\begin{table}[h]
	\centering
	\caption{Phase transition intensity at $f$ values from 500 to 800 GeV.}
	\begin{tabular}{c|c|c|c|c}
		\toprule 
		\hline\hline
		$f$ (GeV)&$\epsilon=v^2/f^2$&$T_c$ (GeV)&$h_m(T_c)$ (GeV)&$S=h_m(T_c)/T_c$\\
		\hline
		\midrule
		500 & 0.24 & 161.69 & 116.22 & 0.7187\\
		530 & 0.22 & 160.59 & 107.21 & 0.6676\\
		560 & 0.19 & 159.72 & 100.90 & 0.6317\\
		590 & 0.17 & 159.03 & 95.50 & 0.6005\\
		620 & 0.16 & 158.45 & 90.99 & 0.5742\\
		650 & 0.14 & 157.98 & 87.39 & 0.5532\\
		680 & 0.13 & 157.58 & 83.78 & 0.5317\\
		710 & 0.12 & 157.23 & 81.08 & 0.5157\\
		740 & 0.11 & 156.94 & 79.28 & 0.5052\\
		770 & 0.10 & 156.68 & 77.48 & 0.4945\\
		800 & 0.09 & 156.46 & 75.68 & 0.4837\\
		\bottomrule
		\hline\hline
	\end{tabular}
	\label{tab:EWPT_CHM_500-800}
\end{table}
\begin{table}[h]
	\centering
	\caption{Phase transition strength at $f$ values from $300$ to $500$ GeV.}
	\begin{tabular}{c|c|c|c|c}
		\toprule
		\hline\hline 
		$f$ (GeV)&$\epsilon=v^2/f^2$&$T_c$ (GeV)&$h_m(T_c)$ (GeV)&$S=h_m(T_c)/T_c$\\
		\hline
		\midrule
		300 & 0.67 & 284.30 & 471.17 & 1.6573\\
		320 & 0.59 & 208.08 & 502.70 & 2.4159\\
		340 & 0.52 & 181.52 & 220.72 & 1.2159\\
		360 & 0.47 & 176.01 & 195.50 & 1.1107\\
		380 & 0.42 & 172.16 & 180.18 & 1.0466\\
		400 & 0.38 & 169.28 & 169.37 & 1.0005\\
		420 & 0.34 & 167.03 & 161.26 & 0.9655\\
		440 & 0.31 & 165.23 & 154.95 & 0.9378\\
		460 & 0.29 & 163.72 & 134.23 & 0.8199\\
		480 & 0.26 & 162.61 & 123.42 & 0.7590\\
		500 & 0.24 & 161.69 & 116.22 & 0.7187\\
		\bottomrule
		\hline\hline
	\end{tabular}
	\label{tab:EWPT_CHM_300-500}
\end{table}
\begin{figure}[H]
	\centering
	\includegraphics[width=0.6\linewidth]{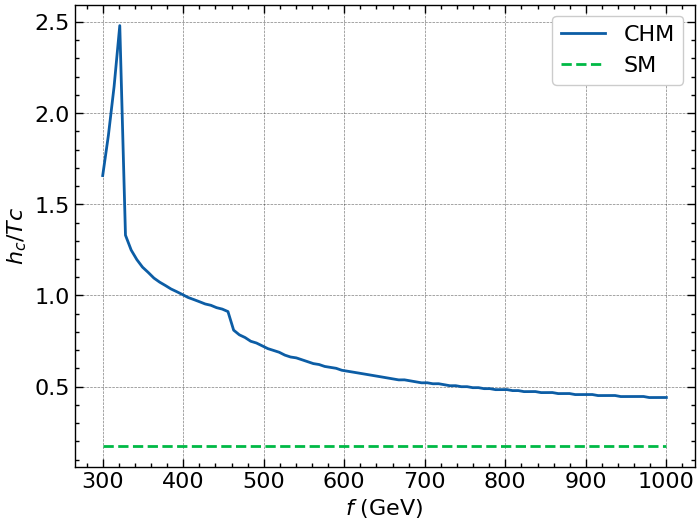}
	\caption{Plot of phase transition intensity in the CHM versus $f$ (purple) and in the SM (blue) in the range from $300$ GeV to $1$ TeV.}
	\label{fig:SMvsCHM_strength_f}
\end{figure}

It can be clearly observed that the phase transition strength in this $f$ region is larger than in the SM case, and the phase transition strength increases as $f$ decreases. Although this strength is still insufficient to trigger a first-order phase transition, the potential of the CHM remains evident seen since the composite part does contribute to the phase transition strength. Indeed, this can be seen more clearly if we consider a larger $f$ region, as long as $f>v$, as shown in \autoref{fig:SMvsCHM_strength_f}. From Fig.\ref{fig:SMvsCHM_strength_f}, it is evident that the phase transition strength continues to increase as $f$ decreases. Referring to the \autoref{tab:EWPT_CHM_300-500}, the phase transition strength has already surpassed 1 when $f<420$ GeV. The graph also exhibits a sharp peak in the $f=300-400$ GeV region. However, the behavior of the graph in the region $\epsilon\approx 1$ is not really meaningful because the effective potential must satisfy $s_h^2\ll1$, and this approximation leads to some VEV Higgs region (not in the entire Higgs VEV domain).

\section{EWPT with dilaton}\label{v}

From the results in the previous section, we can infer that the MCHM is not complete and there may be an alternative way to explain and evaluate the phase transitions of the MCHM. From the beginning of Sec\ref{iv} until now, we have assumed that the composite part breaks the symmetry before the electroweak symmetry breaking occurs. However, this is merely an assumption, because there is no reason why the symmetry of the composite part should be broken before the electroweak breaking. The Higgs-dilaton Composite model is a model built on the premise that symmetries can be broken simultaneously. Specifically, at high energies, the Lagrangian will contain an additional symmetry under the scale transformation $x^\mu\to b^{-1}x^\mu$. Then, this symmetry is broken at the same time as the electroweak breaking, and the breaking is described by the dilaton field $\chi$. The model is now no longer a tunneling of just the Higgs minimum but of the common Higgs-dilaton minimum.

\subsection{The Higgs-dilaton potential in the MCHM at 0K}

There have been numerous analyses and numerical investigations of the dilaton potential and the interactions between the dilaton and the remaining fields in the model \cite{bruggissera, bruggisserb, bruggisserc, von_harling}. Specifically, the dilaton field induces the breaking of the scale symmetry when near VEV $\chi$. Since both the electroweak and scale symmetries are broken at the same time, the symmetry breaking occurs at the same momentum threshold $m_*$.
\begin{equation}
	m_*=g_*f=g_\chi\chi.
\end{equation}
Thus, in the effective potential, $f$ now acts as a dynamic variable, not a constant as in the previous MCHM. To correct the Higgs potential to include the dilaton field, we can fix $f$ to a value consistent with the old MCHM, $f\approx800$ GeV, corresponding to $\chi_0=g_*f/g_\chi$. Then, the zero temperature potential of the Higgs is corrected by replacing the dynamic variable $f$ with $f\chi/\chi_0$ and $f$ is fixed.
\begin{equation}
	V_h(\chi,h)=\qty(\frac{\chi}{\chi_0})^4V_0\qty(\frac{h\chi_0}{\chi})=\qty(\frac{\chi}{\chi_0})^4\qty(\alpha^0\sin^2\frac{h}{f\chi/\chi_0}+\beta^0\sin^4\frac{h}{f\chi/\chi_0}).
\end{equation}
The factor $(\chi/\chi_0)^4$ appears because $\alpha^0$ and $\beta^0$ are approximately proportional to $f^4$. Instead of leaving it as it is, we change the variable $\hat h=h\chi_0/\chi$. 

The simplest way to introduce an invariant kinetic term and still keep the symmetry $\hat h\to \hat h+2\pi k\ (k\in\mathbb Z)$ (because of the Goldstone matrix after switching $f\to f\chi_0/\chi$). By introducing a new dynamical variable $\hat h$, the mixed kinetic term of $h$ and $\chi$ has been included in the kinetic term of $\hat h$. This is clearly seen in Ref.\cite{bruggisserb}. Therefore, the kinetic energy of the Higgs is corrected to
\begin{equation}
	\frac{1}{2}(\partial_\mu\chi)^2+\frac{1}{2}\qty(\frac{\chi}{\chi_0})^2\qty(\partial_\mu\hat h)^2,
\end{equation}
$g_\chi$ and $g_*$ are the effective coupling constants from the large $N$ approximation,
\begin{equation}
	g_*=\frac{4\pi}{\sqrt{N}},\quad g_\chi=\frac{4\pi}{\sqrt{N}}\qty(\text{dilaton meson})\qq{or}g_{\chi}=\frac{4\pi}{N}(\text{dilaton glueball}).
\end{equation}
In which $N$ is the color number in the gauge theory $SU(N)$ at high energies, the dilaton meson will resemble the semi-composite interaction of the fermion part in the CHM, and the dilaton glueball will resemble the interaction of quarks in the technicolor model. Then, $\chi_0$ also has two different values corresponding to the two cases above,
\begin{equation}
	\chi_0=\frac{g_*f}{g_\chi}=f\times\begin{cases}
		1&\qq{for dilaton meson}.\\
		\sqrt{N}/2&\qq{for dilaton glueball}.
	\end{cases}
\end{equation}

Furthermore, since the Yukawa constant (specifically that of top quark) is also formed from the VEV of the dilaton, this constant is also changed by the renormalization equation (RG), specifically the interaction between top quark and the composite part is changed,
\begin{equation}\label{eq:RG_yukawa}
	\beta_\lambda(\chi)\equiv\pdv{\lambda_{t_{L,R}}}{\ln\chi}=\gamma_\lambda\lambda_{t_{L,R}}+\frac{c_\lambda}{g_*^2}\lambda_{t_{L,R}}^3\quad\qty(y_t\approx\frac{\lambda_{t_L}\lambda_{t_R}}{g_*}).
\end{equation}

By the NDA estimation method, the coefficients $\alpha^0$ and $\beta^0$ in the Higgs potential (with the $\vb5\oplus\vb5$ representation of fermion) have the form
\begin{equation}
	\alpha^\text{NDA}=c_\alpha\frac{N_c}{16\pi^2}\lambda_t^2g_*^2f^4,\quad \beta^\text{NDA}=c_\beta\frac{N_c}{16\pi^2}\lambda_t^4f^4\quad(\lambda_t=\max\{\lambda_{t_L},\lambda_{t_R}\}).
\end{equation}
This contradicts $\alpha^0$ and $\beta^0$ because $\alpha^\text{NDA}\gg\beta^\text{NDA}(g_*\gg\lambda_t)$ while we need $\alpha^0\ll\beta^0$. This is a correction problem in the MCHM. In fact, the least possible case that requires correction is the multiplet $q_L$ in the $\vb{14}$ representation of \SO{5} and the singlet $t_R$. This representation still keeps the potential at $V_0$, but $\alpha^\text{NDA}$ and $\beta^\text{NDA}$ are now of the same size,
\begin{equation}
	\alpha^\text{NDA}=c_\alpha\frac{N_c}{16\pi^2}\lambda_t^2g_*^2f^4,\quad \beta^\text{NDA}=c_\beta\frac{N_c}{16\pi^2}\lambda_t^2g_*^2f^4\quad(\lambda_t=\lambda_{t_L}).\label{49}
\end{equation}
The correction required is now much smaller than the $\vb 5\oplus\vb5$ representation. We will use this result to correct the Higgs potential. Specifically, since $\lambda_t$ varies with dilaton, we can correct by
\begin{align}
	\alpha^0&\longrightarrow\alpha^0+\alpha^\text{NDA}(\chi)-\alpha^\text{NDA}(\chi_0),\\
	\beta^0&\longrightarrow\beta^0+\beta^\text{NDA}(\chi)-\beta^\text{NDA}(\chi_0).
\end{align}
Thus, the full form of the Higgs potential at any dilaton $\chi$ has the form
\begin{equation}\label{eq:CHMdilaton_Vhiggs}
	\begin{aligned}[t]
		V_h\qty(\chi,\hat h)&=\qty(\frac{\chi}{\chi_0})^4\qty(V_h^0\qty(\hat h)+V_h^\text{NDA}\qty(\chi,\hat h)).\\
		V_h^0\qty(\hat h)&=\alpha^0s_h^2+\beta^0s_h^4.\\
		V_h^\text{NDA}\qty(\chi,\hat h)&=\qty[\alpha^\text{NDA}(\chi)-\alpha^\text{NDA}(\chi_0)]s_h^2+\qty[\beta^\text{NDA}(\chi)-\beta^\text{NDA}(\chi_0)]s_h^4.\\
		&=\frac{N_c}{16\pi^2}g_*^2f^4\qty[\lambda_t^2(\chi)-\lambda_t^2(\chi_0)]\qty(c_\alpha s_h^2+c_\beta s_h^4).
	\end{aligned}
\end{equation}
$\lambda_t^2(\chi)$ is calculated from the RG equation \eqref{eq:RG_yukawa},
\begin{align}
	&\ln\lambda_t-\frac{1}{2}\partial\ln(1+\frac{c_\lambda}{g_*^2}\lambda^2_t)=\gamma_\lambda\partial\ln\chi,\\
	&\begin{aligned}[b]
		\lambda_t^2(\chi)&=\frac{\gamma_\lambda}{\qty(\dfrac{\chi_0}{\chi})^{2\gamma_\lambda}\qty[\dfrac{\gamma_\lambda}{\lambda^2_t(\chi_0)}+\dfrac{c_\lambda}{g_*^2}]-\dfrac{c_\lambda}{g_*^2}}\\
		&=\frac{\qty(\chi/\chi_0)^{2\gamma_\lambda}}{1/\lambda^2_t(\chi_0)+c_\lambda/(g_*^2\gamma_\lambda)\qty[1-\qty(\chi/\chi_0)^{2\gamma_\lambda}]},
	\end{aligned}
\end{align}
with $\lambda_{t_L}(\chi_0)\propto \sqrt{y_tg_*}$. The Yukawa coupling of top quark, $y_t$, now also depends on $\chi$ and is assumed to have the form,
\begin{equation}
	y_t(\chi)=\frac{\lambda_{t_L}}{g_*}\qty[\lambda_{t_R1}(\chi)+\lambda_{t_R2}],
\end{equation}
with $\lambda_{t_L}$ and $\lambda_{t_R2}$ being the initial parameters, and $\lambda_{t_R1}(\chi)\equiv\lambda_{t_R}(\chi)$.

From the general form of Higgs-dilaton potential (Eq. (\ref{eq:CHMdilaton_Vhiggs})), if $\chi<\chi_0$, the NDA part is dominant, so that $\alpha\approx\alpha^\text{NDA}$ and $\beta\approx\beta^\text{NDA}$. While we require $|\alpha|\ll|\beta|$ to satisfy the EWPT condition $\sin^2(h/f)=\epsilon^2\ll1$ (Eq. (\ref{4})), $\alpha^\text{NDA}$ is of the same order as $\beta^\text{NDA}$ (Eq. (\ref{49})). Hence, although there is a non-zero minimum for $V_h$ because we choose $c_ac_b<0$, the EW symmetry is not broken. Only when $\chi=\chi_0$, $\alpha=\alpha^0\ll\beta=\beta^0$ and $V_\chi$ also reaches the minimum, the two symmetries are broken at the same time.

\subsection{The dilaton potential}

\begin{figure}[h]
	\centering
	\includegraphics[width=0.8\linewidth]{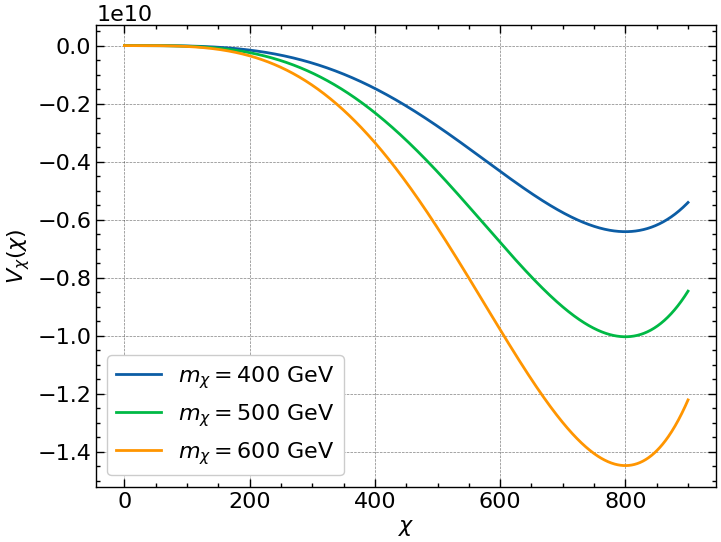}
	\caption{The dilaton potential at the dilaton masses are $400$ GeV, $500$ GeV and $600$ GeV (from top to bottom), respectively).}
	\label{fig:V0_dilaton}
\end{figure}

Before considering the phase transition of the Higgs-dilaton potential, we need to determine the potential of the dilaton itself. The dilaton potential is approximated at zero temperature as
\begin{equation}\label{eq:CHMdilaton_ Vdilaton}
	V_\chi=c_\chi g_\chi^2\chi^4-\varepsilon(\chi)\chi^4,
\end{equation}
with $\varepsilon$ depending on the RG equation,
\begin{equation}\label{eq:RG_dilaton}
	\beta_\varepsilon(\chi)\equiv\pdv{\varepsilon}{\log\mu}=\gamma_\varepsilon\varepsilon+c_\varepsilon\frac{\varepsilon^2 }{g_\chi^2},
\end{equation}
$c_\chi$ and $c_\varepsilon$ are $\order{1}$ coefficients, while $\gamma_\varepsilon$ is assumed to be very small. The first term is the scale invariant, while the second term is responsible for breaking the scale symmetry depending on VEV $\chi$. Solving the above equation at $\mu=g_\chi\chi$, we find
\begin{equation}
	\varepsilon(\chi)=\frac{(\chi/\chi_0)^{\gamma_\varepsilon}}{\dfrac{1}{\varepsilon_0}+\dfrac{c_\varepsilon}{\gamma_\varepsilon g_\chi^2}\qty[1-\qty(\dfrac{\chi}{\chi_0})^{\gamma_\varepsilon}]}.
\end{equation}
$\gamma_\varepsilon$ and $\varepsilon_0\equiv\varepsilon(\chi_0)$ are determined by minimizing the potential at $\chi=\chi_0$ and yielding the correct dilaton mass, a parameter is chosen initially. Minimizing $V_\chi$, we get 
\begin{equation}
	\varepsilon_0[\gamma_\varepsilon]=\frac{8c_\chi g_\chi^2}{(4+\gamma_\varepsilon)+\sqrt{(4+\gamma_\varepsilon)^2+16c_\chi c_\varepsilon}}
\end{equation}
and $\gamma_\varepsilon$ is the solution of the equation,
\begin{equation}
	-\frac{m_\chi^2}{\chi_0^2}=4\qty(c_\chi g_\chi^2-\varepsilon_0[\gamma_\varepsilon])\qty(\gamma_\epsilon+4+c_\varepsilon\frac{2\epsilon_0[\gamma_\varepsilon]}{g_\chi^2}),\label{60}
\end{equation}
with $\epsilon_0\equiv\epsilon(\chi_0)$. The dilaton mass affects the depth of $V_\chi$. It can be seen from \autoref{fig:V0_dilaton} that the larger the dilaton mass is, the faster the dilaton potential decreases and the lower its minimum becomes.

\subsection{The high-temperature potential}
Having determined the dilaton's own potential and the Higgs-dilaton potential, we can now move on to studying the potential at finite temperature. At $\chi=0$, the normalized potential is
\begin{equation}
	V^\beta_\text{eff}(0)=-\frac{\pi^2}{8}N^2T^4-\frac{\pi^2g_\text{SM}}{90}T^4,
\end{equation}
where $g_\text{SM}\approx100$ is the total number of degrees of freedom of the entire SM. The first term is the value from the super Yang Mills theory \SU{$N$} with $\mathcal N=4$. When considering the dilaton potential at finite temperature, the dilaton interval is divided into two regions: (I) $0\lesssim\chi\lesssim T/g_\chi$ and (II) $\chi\gtrsim T/g_\chi$. In region (II), the strong confinement scale $g_\chi\chi$ is larger than the temperature under consideration, so the contribution of composite fields to the effective potential is very small. Therefore, the potential at a finite temperature also consists only of the corrected Higgs potential and the contributions of SM particles to the dilaton potential:
\begin{equation}
	V^\beta_\text{eff}(\chi\gtrsim T/g_\chi, \hat h)=V_h+V_\chi+\Delta V_\text{SM}^\beta,
\end{equation}
where $\Delta V_\text{SM}^{\beta}$ denotes the temperature contributions of the SM particles, including Higgs, which are calculated using the $J_{B,F}$ functions.

In region (I), the temperature is now higher than the strong confinement scale, so we need to know the full description in the ultraviolet limit to calculate the temperature-dependent corrections accurately. Since the phase transition does not occur in this region either, the simplest way is to "connect" the potential at $\chi=0$ and region (II) by a step function.

Temporarily ignoring the Higgs-dependent terms and the contributions at finite temperature, we calculate the dilaton transition temperature by solving
\begin{equation}
	V^{1/T_c}_\text{eff}(0)=V^{1/T_c}_\text{eff}(\chi_\text{min}\gtrsim T_c/g_\chi).
\end{equation}

We also approximate the large $N$ strongly interacting color limit to ignore the SM term in $V^\beta_\text{eff}(0)$,
\begin{equation}
	\begin{aligned}[b]
		-\frac{\pi^2}{8}N^2T_c^4\approx V_\chi(\chi_0)&=\qty(1-\frac{8}{4+\gamma_\varepsilon+\sqrt{(4+\gamma_\varepsilon)^2-16c_\varepsilon c_\chi}})c_\chi g_\chi^2\chi_0^4\\
		&\approx\qty(1-\frac{1}{1+\gamma_\varepsilon/4})c_\chi g_\chi^2\chi_0^4\approx\frac{\gamma_\varepsilon}{4}c_\chi g_\chi^2\chi_0^4=\frac{\gamma_\varepsilon}{4}\frac{g_*^4}{g_\chi^2}c_\chi f^4.
	\end{aligned}
\end{equation}
Thus the phase transition temperature can be estimated as
\begin{equation}
	T_c\approx\frac{g_*f}{\sqrt{\pi N g_\chi}}\qty(-2\gamma_\varepsilon c_\chi)^{1/4}=2f\qty(-2\gamma_\varepsilon c_\chi)^{1/4}\times\begin{cases}
		N^{-3/4}&\text{for dilaton meson},\\
		(2N)^{-1/2}&\text{for dilaton glueball}.
	\end{cases}
\end{equation}

This means that the phase transition strength increases with increasing $N$. Furthermore, considering the components that we have approximately neglected would deepen the potential at the minimum and thus reduce the transition temperature.

However, instead of using a step function, it is more convenient to adjust the effective potential to be a smooth and continuous function over region (I). We will enter a term in the effective potential in region (II) that is the one-loop temperature contribution of the CFTs, $\Delta V_\chi^\beta(\chi,T)$,
\begin{equation}
	V_\text{eff}(\chi,\hat h,T)=V_h+V_\chi+\Delta V^\beta_\text{SM}+\Delta V^\beta_\chi(\chi,T),
\end{equation}
with
\begin{equation}
	\Delta V^\beta_\chi\qty(\chi,T)=\sum_\text{CFT boson}\frac{n_BJ_B\qty(\beta^2 g_\chi^2\chi^2)}{2\pi^2\beta^4}-\sum_\text{CFT fermion}\frac{n_FJ_F\qty(\beta^2 g_\chi^2\chi^2)}{2\pi^2\beta^4}.\label{67}
\end{equation}
Since the above two contributions are not much different, we can assume that only the fermion resonance contribution is zero and
\begin{equation}
	\sum_\text{boson resonance}n=\frac{45N^2}{4}
\end{equation}
so that $V_\text{eff}\qty(\chi,\hat h,T)$ returns the same value in the super Yang-Mills model as in the remaining domain.

\subsection{The numeritical results}

First, we can estimate that the electroweak phase transition with the addition of the dilaton potential is a first-order phase transition, due to the contribution of Eq. \eqref{67}. The Higgs-dilaton potential also has many unknown parameters. Therefore, finding a suitable parameter domain is a necessary step. In Ref. \cite{bruggissera}, seven parameters are selected as shown in Table \ref{tab:EWPT dilaton parameter}. The choice of values of the parameters may differ from Table \ref{tab:EWPT dilaton parameter}; however, this is a complex choice but their values may fluctuate around the values in Table \ref{tab:EWPT dilaton parameter}.

\begin{table}[h]
	\centering
	\caption{Parameters involved in solving the Higgs-dilaton EWPT.}
	\begin{tabular}{|c|c|c|c|c|c|c|}
		\hline
		$c_\varepsilon$&$c_\chi$&$c_\alpha$&$c_\beta$&$\gamma_{\lambda}$&$c_\lambda$&$\lambda_{t_R}(\chi_0)$\\
		\hline
		0.5&0.2&$-0.3$&0.3&$-0.3$&1.875&$0.6\sqrt{y_tg_*}$\\
		\hline
	\end{tabular}
	\label{tab:EWPT dilaton parameter}
\end{table}
We conduct the EWPT survey with the Higgs-dilaton potential as follows:
\begin{itemize}
	\item\textbf{Step 1}: Choose the parameters as in the \autoref{tab:EWPT dilaton parameter} (in Ref. \cite{bruggissera}), $f=800$ GeV. Considering the case where the dilaton is a meson, we can deduce that $\chi_0=800$ GeV. Determine $\lambda_t(\chi)$ through the RG Eq.\eqref{eq:RG_yukawa}.
	\item\textbf{Step 2}: Choose a value of $m_\chi$ to find $\varepsilon(\chi)$ by solving the system of differential equations:
	\begin{equation}
		\begin{cases}
			\text{The RG Equation }\eqref{eq:RG_dilaton},\\
			V_\chi'(\chi_0)=0,\\
			V_\chi''(\chi_0)= m_\chi^2.
		\end{cases}
	\end{equation}
	This gives the specific form of the dilaton potential Eq.\eqref{eq:CHMdilaton_ Vdilaton}. For the larger $m_\chi$ is, the larger $\varepsilon$ becomes in the range $\chi$ from 0 to $\chi_0$ (\autoref{fig:epsilon_dilaton}).
	\begin{figure}[H]
		\centering
		\includegraphics[width=0.8\linewidth]{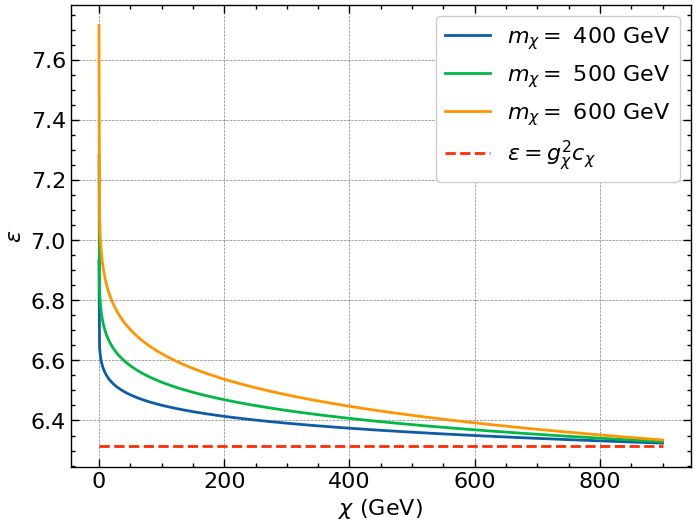}
		\caption{$\epsilon(\chi)$ with different dilaton masses.}
		\label{fig:epsilon_dilaton}
	\end{figure}
	\item\textbf{Step 3}: Determine the remaining potential terms in the total potential.
	\begin{itemize}
		\item The potential $V_h$ according to Eq. \eqref{eq:CHMdilaton_Vhiggs}.
		\item The one-loop contributions according to the Coleman-Weinberg formula with mass functions recalibrated to contain $\chi$,
		\begin{align}
			&M^2_{W,Z}\qty(\chi, \hat h)=\qty(\frac{\chi}{\chi_0})^2M^2_{W,Z}\qty(\hat h)=\qty(\frac{\chi}{\chi_0})^2\frac{m_{W,Z}^2}{\epsilon}\sin^2\frac{\hat h}{f}.\label{70}\\
			&M_t^2\qty(\chi,\hat h)=\qty(\frac{\chi}{\chi_0})^2\frac{y_t^2(\chi)}{2}f^2\sin^2\frac{\hat h}{f}.\label{71}\\
			&M_h^2(\chi,\hat h)=\qty(\frac{\chi}{\chi_0})^2\frac{1}{f^2}\qty[2\alpha(\chi)(c_h^2-s_h^2)+\beta(\chi)(12s_h^2c_h^2-4s_h^4)].\label{72}
		\end{align}
		\item The one-loop contributions using Coleman-Weinberg formula after renormalization,
		\begin{equation}
			\Delta V^1_h(\chi,\hat h)=\qty(\frac{\chi}{\chi_0})^4\sum_{i=h,W,Z,t}\qty[M^4_i\qty(\log\frac{M^2}{m^2}-\frac{3}{2})+2M_i^2m^2_i],
		\end{equation}
		with $m_i=M_i(\chi,\ev{h})$.
		\item The one-loop contributions of $h,W,Z,t$ at temperatures, to the higgs potential by the temperature functions $J_B$ and $J_F$
		\begin{equation}
			\Delta V_\text{SM}^\beta\qty(T,\chi,\hat h)=\sum_{i=h,W,Z}\frac{n_iT^4}{2\pi^2}J_B\qty(\frac{M_i^2}{T^2})-\frac{12T^4}{2\pi^2}J_F\qty(\frac{M_t^2}{T^2}).
		\end{equation}
	\end{itemize}
And the total potential is
	\begin{equation}
		V_\text{eff}(T, \chi, \hat h)=V_\chi(\chi)+V_h\qty(\chi,\hat h)+\Delta V^1_h\qty(\chi,\hat h)+\Delta V_\text{SM}^\beta\qty(T, \chi,\hat h)+\Delta V^\beta_\chi\qty(\chi,T).\label{75}
	\end{equation}
Note that the total potential needs to be normalized so that $V_\text{eff}(T,0,0)=0$.

When we switch $f\to f\chi/\chi_0$, the dilaton have already involved in the one-loop contributions to the Higgs potential through the new mass functions in Eqs. (\ref{70}-\ref{72}). That is, at the one-loop level, we consider the contributions of the SM particles with the mass components corrected by dilaton (as seen in Eqs. (\ref{70}-\ref{72})). Therefore, dilaton still has a thermal contribution to the potential (Eq. \ref{75}). Furthermore, the thermal contribution of the dilaton is also expressed through $\Delta V^\beta_\chi(\chi,T)$ (Eq. (\ref{67})).

In addition, we do not consider the thermal contribution of the dilaton alone, because the mass component of the dilaton (as seen in Eq.(\ref{60})) comes from the dilaton potential, which does not interact with the Higgs field (although dilaton interacts with the Higgs field and only to the 4th order as in Eq.(\ref{eq:CHMdilaton_Vhiggs}), the mass component of the dilaton does not come from interacting with the Higgs field). Again, we consider the thermal contributions of SM particles at the one-loop level, because their mass comes from interacting directly with the Higgs field. However, at higher orders, the higher-order thermal contributions of the dilaton must still be considered.

We consider the full one-loop effective potential, when considering the quantum correction component of the particles at zero temperature, to make the numerical solution converge better. The $F$ functions are quite difficult to calculate because of the weak convergence.

	\item\textbf{Step 4}: Find the critical temperature $T_c$ and $\qty(\chi_c,\hat h_c)$ which are the minima of $V_\text{eff}(T_c, \chi_c,\hat h_c)$ such that
	\begin{gather}
		V_\text{eff}(T_c,\chi_c,\hat h_c)=-\frac{\pi^2N^2}{8}T_c^4-\frac{\pi^2g_{SM}}{90}T^4_c,\\
		\intertext{or}
		V_\text{eff}(T_c,\chi_c,\hat h_c)+\frac{\pi^2N^2}{8}T_c^4+\frac{\pi^2g_{SM}}{90}T^4_c=0.\label{eq:numerical2}
	\end{gather}
Then $T_c$ is the phase transition temperature. Fixing the CFT color number $N=5$ and $g_\text{SM}=100$.
\end{itemize}

We find that the two most important parameters are $N$ and $m_{\chi}$, which is consistent with the surveys in Refs. \cite{bruggissera,bruggisserb,bruggisserc}. $T_c$ is calculated while ignoring the thermal contribution of the SM particles, showing that it increases as $N$ increases. So we choose a value of $N$ as a benchmark. According to Refs. \cite{bruggisserb,bruggisserc}, $N$ in the range of 4 to 7 will give good results on EWPT. $N=5$, or near 5, will give the largest $\omega_{tot}$ (the total washout factor of the baryon asymmetry). So we choose $N=5$.

The other parameters are chosen similarly to Refs. \cite{bruggissera,bruggisserb,bruggisserc}. But the first thing we want to do is to examine the EWPT with the same parameters as in the references for verification. We see that, from the results, this set of parameters is suitable. In other words, we have filtered the optimal parameters from Refs. \cite{bruggisserb,bruggisserc}.

We have given a specific procedure to calculate the phase transition intensity. This procedure can be used with different sets of parameters. We also assume $m_{\chi}$ in the interval [300,800] GeV, and in Refs. \cite{bruggissera,bruggisserb,bruggisserc} $m_{\chi}$ is in the interval [300,600] GeV, and also in these intervals the first-order electroweak phase transitions exist. So our surveys are compatible with Refs. \cite{bruggisserb,bruggisserc}, but our domain of $m_{\chi}$ is wider than in Refs. \cite{bruggisserb,bruggisserc}.

Another important thing is that we want to present a picture going from the CHM without dilaton to the need to add dilaton into the EWPT problem. Although initially the dilaton was introduced due to other physical reasons. Or our result shows that if $f$ is small enough, then the EWPT is a first-order phase transition, but it is not physical, so it is necessary to add dilaton. Therefore, our filtering of parameters to investigate the EWPT is appropriate and seamless with the investigations in Refs. \cite{bruggisserb,bruggisserc}.

\begin{figure}[H]
	\centering
	\includegraphics[width=0.85\linewidth]{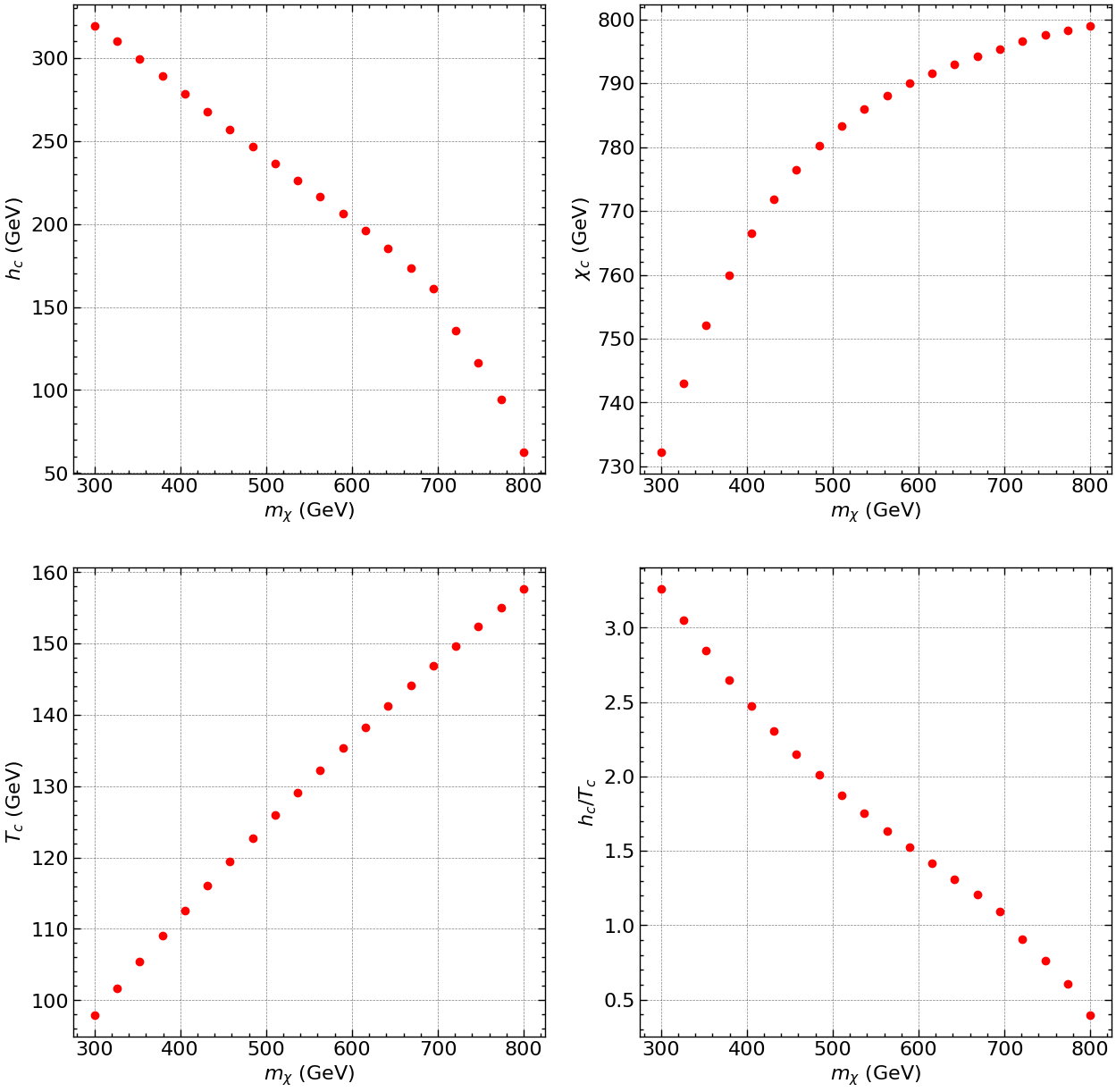}
	\caption{Phase transitions with dilaton masses in the range from $300$ GeV to $800$ GeV.}
	\label{fig:EWPT_higgs_dilaton}
\end{figure}

The above steps are performed with $m_\chi$ in the mass range from 300 to 800 GeV. The results give four graphs corresponding to $\chi_c$, $h_c$, $T_c$ and $h_c/T_c$ against the dilaton mass $m_\chi$ in \autoref{fig:EWPT_higgs_dilaton} and the corresponding data at \autoref{tab:data_CHM_dilaton_300-1000}. It can be seen from the graphs that the dilaton mass clearly affects the minimum state of the model. As the dilaton mass increases, $\chi_c$ also increases and gets closer and closer to $\chi_0$. This shows that the larger the dilaton mass is, the less the minimum of the total potential energy is affected by the Higgs potential, leading to $\chi_0$ also being almost the minimum of the total potential. This is also reasonable because the minimum of the dilaton potential becomes deeper and narrower. Also, the required transition temperature increases with the depth of the minimum, and as the temperature increases, the Higgs potential minimum becomes increasingly smaller due to the contribution at a finite temperature.
\begin{table}[h]
	\centering
	\caption{Phase transitions at $f=800$ GeV and $m_\chi$ from 300 GeV to 800 GeV.}
	\begin{tabular}{c|c|c|c|c}
		\toprule 
		\hline\hline
		$m_\chi$ (GeV)&$T_c$ (GeV)&$h_c(T_c)$ (GeV)&$\chi_c(T_c)$ (GeV)&$S=h_c(T_c)/T_c$\\
		\hline
		\midrule
		300.00 & 97.92 & 732.21 & 319.43 & 3.2623\\
		326.32 & 101.67 & 742.92 & 309.84 & 3.0475\\
		352.63 & 105.37 & 752.12 & 299.56 & 2.8430\\
		378.95 & 109.00 & 759.92 & 288.91 & 2.6506\\
		405.26 & 112.55 & 766.45 & 278.15 & 2.4713\\
		431.58 & 116.03 & 771.89 & 267.45 & 2.3051\\
		457.89 & 119.42 & 776.41 & 256.91 & 2.1513\\
		484.21 & 122.74 & 780.17 & 246.56 & 2.0087\\
		510.53 & 125.99 & 783.30 & 236.38 & 1.8762\\
		536.84 & 129.16 & 785.93 & 226.31 & 1.7521\\
		563.16 & 132.27 & 788.14 & 216.26 & 1.6350\\
		589.47 & 135.31 & 790.01 & 206.14 & 1.5235\\
		615.79 & 138.29 & 791.62 & 195.80 & 1.4158\\
		642.11 & 141.22 & 793.00 & 185.06 & 1.3104\\
		668.42 & 144.09 & 794.21 & 173.62 & 1.2050\\
		694.74 & 146.90 & 795.30 & 160.87 & 1.0950\\
		721.05 & 149.67 & 796.65 & 135.70 & 0.9066\\
		747.37 & 152.40 & 797.54 & 116.48 & 0.7643\\
		773.68 & 155.08 & 798.32 & 94.44 & 0.6090\\
		800.00 & 157.72 & 799.05 & 62.62 & 0.3970\\
		\hline\hline
		\bottomrule
	\end{tabular}
	\label{tab:data_CHM_dilaton_300-1000}
\end{table}

Considering the possibility of first-order transition, the overall transition strength in the Higgs-dilaton model is larger than in both the SM and CHM in the range of $m_\chi$ considered. The transition strength can be larger than 1 when $m_\chi<700$ GeV. Furthermore, the minimum of the dilaton potential lies in the vicinity of $\chi_0$, while $f$ in the CHM is only below 400 GeV for $h_c/T_c>1$. This shows that the dilaton not only helps to trigger the first-order EWPT but also helps this process to occur in a suitable value range, without breaking the approximation of $\sin^2(h/f)$.

\section{CONCLUSION AND OUTLOOKS}\label{vi}
	
The paper shows that the Minimal Composite Higgs Model has a stronger electroweak transition strength than the Standard Model. The SM is most likely an effective theory and the effect of the composite components changes the shape of the potential, leading to a change in the electroweak transition.

The MCHM alone is not sufficient to trigger an electroweak transition compatible with the conditions of $f$. When $f$ is varied in the range $300-800$ GeV the model suggests that with a confinement scale $f$ smaller than a suitable value, the transition strength can be achieved. This leads to the possibility that $f$ is a dynamical field and the electroweak transition occurs before $f$ reaches its vacuum value. Because if $f$ varies as a dynamic variable in the effective potential, it can establish a strong first-order phase transition, as examined in Section \ref{iv}. However, $f$ is a constant, with values greater than $300$ GeV. Therefore, it has been suggested to introduce a dilaton field that acts as a dynamical field instead of $f$ in the electroweak phase transition problem. The fixed confinement scale $f$ is then upgraded to a dilaton dynamical field $\chi$. By modifying the Composite Higgs model to include the dilaton appropriately, the barrier between the two vacuums is significantly increased, leading to a naturally large phase transition strength of 1 and the electroweak phase transition still occurring in the dilaton region consistent with the approximations in the Composite Higgs model. At the zero temperature, the dilaton potential and the Higgs potential remain independent of each other in Eq. \eqref{eq:numerical2}. However, this independence may not remain if they all come from the same composite sector. And at finite temperatures the Higgs and dilaton interact with each other through the presence of the dilaton in the Higgs potential, thereby triggering the first-order EWPT.

Physically, the presence of the dilaton may be an important contribution of the composite part that we may have overlooked when making the approximations in the MCHM. This occurs in the CHM due to the complexity of translating high-energy interactions to low-energy ones (the $\Pi$ functions in the interaction of top quark and the resonance fields) and the lack of understanding of physics at very high energies to be able to give the correct form of the Lagrangian as well as the effective potential.

The study of the Higgs-dilaton model in the paper has shown the potential of the dilaton in satisfying the thermodynamic condition of one of the three Sakharov conditions, however, this model still has many unknown parameters, because the knowledge of physics at very high energies is still limited. Therefore, the results require investigation with many sets of parameters. In addition, because the dilaton increases the complexity of the potential as well as the number of variables of the system, the calculation needs to be done numerically and requires a lot of resources to be able to investigate with numerous different parameter cases.

Based on the techniques in this paper, our next work is to compute the Sphaleron energy, recheck the decoupling condition, and reconfirm $S>1$. With the introduction of the dilaton, further studies on the effects of the dilaton on the interaction constants and experimentally tested parameters can be conducted to constrain the possible parameters. In addition, the results can be further constrained by further studies on the effects of the dilaton on the remaining conditions for baryogenesis from the electroweak phase transitions.
	
\section*{ACKNOWLEDGMENTS}
This research is funded by Vietnam National Foundation for Science and Technology Development (NAFOSTED) under grant number 103.01-2023.16.

\end{document}